\documentclass[prl,nofootinbib,showpacs,preprint,tikz]{revtex4}
\usepackage{lmodern} 
\usepackage{esvect} 
\usepackage{tensor}
\usepackage{bm}
\newcommand{\al}{\alpha}
\usepackage{numprint}
\usepackage{tikz}
\usetikzlibrary{shapes}
\usepackage{pgf}
\usetikzlibrary{arrows}

\usepackage{subfigure}
\usepackage{amsmath}
\usepackage{amsfonts}
\usepackage{amssymb}
\usepackage{graphicx}       

\usepackage[colorlinks]{hyperref}

\usepackage[applemac]{inputenc}
\usepackage[T1]{fontenc}

\usepackage{bbm}
\def\beq{\begin{equation}}
\def\eeq{\end{equation}}
\def\bea{\begin{eqnarray}}
\def\eea{\end{eqnarray}}
\def\ba{\begin{array}}
\def\ea{\end{array}}
\def\nn{\nonumber \\}

\def\ie{{\it i.e. }}
\def\lb{\left( }
\def\rb{\right) }
\def\ame{&=&}
\def\am+{&+&}
\def\nn{\nonumber\\}

\def\part{\partial}
\def\tfrac#1#2{{\textstyle{#1\over #2}}}

\def\haf#1{\tfrac{#1}{2}}
\def\x{\times}

\newtheorem{theorem}{Theorem}[section]

\begin{document}

\def\w{\par \vspace{\baselineskip}}

\preprint{UdeM-GPP-TH-21-288 }
\preprint{arXiv:1208.2293[hep-th]}
\title{Near Zone Dynamical Effects in Gravity }
\author{Victor Massart$^{1}$}
\email{victor.massart@umontreal.ca}
\author{M. B. Paranjape$^{1,2}$}
\email{paranj@lps.umontreal.ca}

\affiliation{$^1$Groupe de physique des particules, D\'epartement de
physique and $^2$Centre de recherche mathŽmatiques, Universit\'e de Montr\'eal, C.P. 6128, succ. centre-ville,
Montr\'eal, Qu\'ebec, CANADA, H3C 3J7 }


\begin{abstract}
Dynamical effects in general relativity have  been finally, relatively recently observed by  LIGO\cite{2016LRR....19....1A}.  These effects correspond to gravitational waves created by the coalescence of black holes or neutron stars billions of years ago and and billions of light years away from their sources.  To be able to measure these signals, great care has to be taken to minimize all sources of noise in the detector.  One of the sources of noise is called Newtonian noise, the name based on the notion that close by sources would create essentially instantaneous Newtonian gravitational fields.   In this article we present an analysis of the dynamical (time dependent) nature of the Newtonian noise.  In that respect, it is a misnomer to call it Newtonian noise, the Newtonian theory does not afford any dynamical notion of the gravitational field.  We will in fact do our analysis in the context of Einsteinian general relativity.  The dynamical aspects of the nature of the Newtonian noise have heretofore been disregarded as they were considered negligible.  However, we demonstrate that they are indeed not far from the realm of being measurable.  They could be used to validate Einsteinian general relativity or to give valuable information on the true dynamical nature of gravity.  One fundamental question, for example,  is a direct measurement the speed of propagation of gravitational effects and the verification that it is indeed the same as the speed of light.  We propose a simple laboratory experiment that could affirm or deny this proposition.  We also analyze the possibility of the detection of large geophysical events, such as earthquakes.   We find that large seismic events seem to be easily observable with the present ensemble of gravitational wave detectors,.  The ensemble of gravitational wave detectors could easily serve as a system of early warning for otherwise catastrophic seismic events.  
\end{abstract}
\maketitle

\tableofcontents

\section{Introduction}

In the quest for observing dynamical gravitational signals, Newtonian noise, often called gravity gradient noise, originating from seismic gravitational disturbances will give the ultimate noise threshold, beyond which no signals could ever be observed \cite{PhysRevD.58.122002}.  Newtonian noise of anthropogenic origin or of other controllable origin has heretofore been analyzed \cite{Thorne:1998hq}, however the focus has been to eliminate this source of noise so that astronomically sourced gravitational waves could be observed.  On the other hand, it is clearly imaginable that we could try to create strong enough and observable Newtonian signals and measure their time dependent, dynamical properties.  Such measurements could give rise to a stunning verification or refutation of Einsteinian general relativity. 

This paper is the continuation of an analysis \cite{ParanjapeSpeedGravity} which gave rise to a computation and a proposal for measuring the speed of gravity, which we will call $c_g$,  in the near-field zone, in a laboratory setting where all aspects are under direct control.  The idea enunciated in \cite{ParanjapeSpeedGravity} observed that a finite propagation speed could give rise to measurable relative aberration of the effects of gravity on a detector and subsequently the ability to measure that speed.  Only recently the best limit on the speed of gravity was set in conjunction with the simultaneous observation of the arrival times of gravitational waves by LIGO\cite{LIGOSpeedGravity} and of gamma rays \cite{PhysRevLett.119.161102} from the same source.  It was found that the speed of gravity, $c_g$, and the speed of light, $c$, were identical to one part in $10^{15}$.   However, it should be noted that these observations are not done in a controlled environment and rely on the assumption that both signals were emitted at the same time.   They are in fact an indirect measurement of the speed of gravity.  They are also observations  in the radiation zone, a totally different regime from the proposed measurement \cite{ParanjapeSpeedGravity} here,  which is in the near zone.   As expounded upon by Saulson \cite{Saulson:2017jlf}, the creation of gravitational waves in the laboratory, where we could control the amplitude, frequency and polarization would make possible an  unequivocal test of general relativity, akin to Heinrich Hertz's experiments which verified Maxwell's prediction of electromagnetic waves.  It would be important to create a gravitational disturbance and measure its arrival at a spatially distanced detector all under the scrutiny of direct, controlled, laboratory experiment.

In the next section we will discuss the theoretical background. First we will discuss  the weak field limit of the Einstein equations\cite{Carroll,Misner} underlining the connections with the equations of electrodynamics.  Then we will discuss the conservation of momentum and  analyze its implications for human created gravity gradient noise\cite{Thorne:1998hq}.  In the following section we will apply these considerations to the question of measuring the velocity of the propagation of gravitation.  We will use the Lagrange inversion theorem which will be exposed and explained (briefly).   Finally we will discuss the possibility for the observation of the calculated effects including the gravity gradient signal that could be created by large enough earthquakes\cite{bc,Bletery1027}.

\section{Theoretical background}

\subsection{Weak field gravitation}
%
We begin with the full Einstein equations,
\begin{equation}
    R_{\mu \nu} - \frac{1}{2} g_{\mu \nu}R = 8 \pi G T_{\mu \nu}.\label{ee}
\end{equation}
We are interested in the weak-field approximation, which corresponds to a restriction to coordinate systems in which we can write the expansion of the metric, $g_{\mu \nu}$, around a (Minkowski) background, as $\eta_{\mu \nu}$ plus a small perturbation, $h_{\mu \nu} \ll 1$\footnote{$\eta_{\mu \nu}$ (diagonal) and our signature are chosen as $(-1, 1, 1, 1)$}. Then we can compute the connections (the Christoffel symbols), the Riemann tensor, the Ricci tensor and curvature scalar, while neglecting at each step the terms $o(h^2)$ and higher.  We do not record the corresponding expressions here, they are well known, see for example \cite{Misner}.  We also make the harmonic gauge choice,
\begin{equation}
    \partial^{\mu} (h_{\mu \nu} - \frac{1}{2}\eta_{\mu \nu} h_{\lambda}^{\lambda}) = \partial^{\mu} \bar{h}_{\mu \nu} = 0.\label{gaugechoice}
\end{equation}
Expressing all quantities in terms of $\bar{h}_{\mu \nu}$ simplifies the notation considerably.

In the harmonic gauge, the  linearized approximation to the Einstein equations \eqref{ee} gives simply
\begin{equation}
    \square \bar{h}^{\mu \nu} = -16\pi G T^{\mu \nu}.
    \label{LinearizedEQ}
\end{equation}
We note that $\bar h^{\mu\nu}$ is dimensionless which requires that $G\to G/c_g^4$, however powers of $c_g$ are suppressed in most equations that follow.  The set of equations in Eqn.\eqref{LinearizedEQ} describe dynamical gravitational phenomena and are in fact very similar to the equations for the electromagnetic potentials in Lorenz gauge, $\square A^{\nu} = 4 \pi j^{\nu}$.
This  similarity is very useful since the dynamics of electromagnetism is well understood, in particular we know the physical (retarded) solution, 
\beq
    \bar{h}^{\mu \nu} (x) =  16\pi G \int D_r(x - x') T^{\mu \nu}(x') d^4x'.
    \label{GravitoPotential}
\eeq
where $D_r(x - x') $ is the retarded Green function of the d'Alembertian. 


We will consider point sources giving rise to gravitational phenomena.  The energy momentum tensor of a gravitational point source is well understood, \cite{Weinberg, Ryder}, and can be written as
\begin{equation}
    T^{\mu \nu} = \frac{M}{\sqrt{1 - \beta^2}} \beta^{\mu} \beta^{\nu}\delta^3(\bm x-\bm{r}(t) )=M\gamma \beta^{\mu} \beta^{\nu}\delta^3(\bm x-\bm{r}(t) )
\end{equation}
where $\bm r(t])$ is the position of the point source, $\beta^{\mu} = (1, \bm\beta)=(1,\frac{1}{c_g }\frac{d}{dt} \bm{r}(t) )$ is its four velocity.  $M$ contains a suppressed factor of $c_g^2$,  $M\to Mc_g^2$.   Corrections to the gravitational field from the non-point like nature of the sources will involve the higher multipoles and will be assumed to be negligible.  The solutions to \eqref{GravitoPotential} are well known, and in electrodynamics are called the Lienard-Wiechert potentials\cite{Jackson:1998nia}.  Correspondingly, the gravitational fields $  \bar{h}^{\mu \nu} (x) $ are then given by
\beq
\bar{h}^{\mu \nu}  (\bm{x}, t) = \left. \left( \frac{4GM\gamma\beta^\mu\beta^\nu}{(1 - \bm{\beta}\cdot \bm{n})R} \right) \right|_{ret} ,\label{ge}
\eeq
where $\bm{R}(t) =  \bm{x} - \bm{r}(t) $, $R(t)=|\bm{R}(t)|$ and $\bm n(t)=\bm R(t)/R(t)$.  $\bm{R}(t)$ is the vector pointing from the (point) source at $\bm{r}(t)$ to the observer at $\bm x$.  The subscript $\scriptstyle{ret}$ means evaluated at the retarded time $t_r$, \ie $\beta(t)\to\beta(t_r)$ etc.  The retarded time $t_r$ is the time at which a source must emit a signal so that it reaches an observer at a given time $t$, explicitly
\beq
t=t_r+R(t_r)/c_g.
\eeq
    $c_g$ is the speed of propagation of gravitational effects, surely the same as the speed of light $c$, however, one of the points of this paper is that this should be experimentally measured and confirmed.  

\subsection{Effective gravitational force}
The expression for the effective force on test bodies results from the geodesic equation  
\begin{equation}
	\frac{dp^\mu}{dt} = - \Gamma^{\mu}_{\rho \sigma} \frac{p^\rho p^\sigma}{E} \ \ \text{with} \ \ p^{\mu} = E V^{\mu}.
\end{equation}
The force can be expanded to second order in the source velocities and we will also expand it to zeroth order in the body velocity  \ie, we will neglect any velocity dependent forces (effective magnetic type forces).   For a body with momentum $\bm{p}=m\bm{V}$ \cite{Carroll},  we find (note that all temporal derivatives come with an unwritten factor of $1/c_g$) 
\begin{align}
	\frac{dp^i}{dt} &= - E \Gamma^i_{00} = -  m \left[ - \frac{1}{2}\partial_i h_{00} + \partial_0 h_{0i} \right]+ \mathcal{O}(V)\nonumber \\
	&= - m \left[ - \frac{1}{4}\partial_i \bar{h}_{00} + \partial_0 \bar{h}_{0i} - \frac{1}{4}\partial_i \bar{h}^j_j \right]+ \mathcal{O}(V)\label{ff}
\end{align}

We note here that the derivatives which appear in \eqref{ff} are with respect to $x^\mu$ while the gravitational fields are functions of the retarded time $t_r$,  as in \eqref{ge} and hence for both temporal and spatial derivatives one has to take this change of variables into account.

Then for a given motion of a point source, we find
\bea
\frac{d\bm p}{dt}\ame
     -Gm M \left[\gamma \frac{\bm{n} - \bm{\beta}}{\kappa^2 R^2} - \frac{\bm{R}}{\kappa^3 R^3} (\dot{R} + \beta^2 - \bm{R}\cdot \dot{\bm{\beta}})\right.\nn 
    \am+ \left.\frac{4\bm{\beta}}{ R^2} \left( \dot{R} + \beta^2 - \bm R \cdot \dot{\bm{\beta}} \right)
     - \frac{4\dot{\bm{\beta}}}{ R} + \frac{\bm{n}}{ R^2} \beta^2 \right]_{ret}
   \label{EffectiveGravitoField11}
\eea
where $\kappa=1-\bm{\beta} \cdot \bm{n}$ and it is understood that only terms up to $\beta^2$ and $\dot{\bm\beta}$ up to ${1}/{c_g^2}$ should be kept inside the bracket.  This gravitational field can be created by moving macroscopic sized masses in the neighbourhood of a detector, such as the mirror in the LIGO experiment \cite{2016LRR....19....1A} and it does seem likely that dynamical predictions of Einsteinian general relativity could be measured.

Examining this formula a little critically, if the motion is uniform, \ie $\dot\beta=0$,  we find
\beq
\frac{d\bm p}{dt}=
     -Gm M \left[ \frac{\gamma}{\kappa^2 R^2} \left( 1 -  \frac{1}{2} (3(\bm n \cdot \bm \beta)^2 -  \beta^2) \right)\bm\eta -  \lb\frac{2 \bm n \cdot \bm \beta }{ R^2}\rb\bm\beta  + \mathcal{O} (\beta^3) \right]_{ret}\label{EffectiveGravitoField2}
\eeq
where the direction of $\bm\eta$ is the direction of the retarded position quadratically extrapolated to the instantaneous direction,  given by
\begin{align}
	\bm\eta &= \bm n + R \frac{d \bm n}{dt} + \frac{1}{2}R^2 \frac{d^2 \bm n}{dt^2}\nn
	&= \frac{1}{\kappa} \left[ \left(1+\frac{1}{2}( 3(\bm n \cdot \bm \beta)^2 - \beta^2 ) \right) \bm n - \left(1 + \bm n \cdot \bm \beta \right) \bm\beta + \mathcal{O} (\beta^3) \right]
\end{align}
and again it is understood that $\kappa$ should be expanded to order $\beta^2$.
We wish to emphasize that  formula in Eqn.\eqref{EffectiveGravitoField2} is strictly valid for unaccelerated motion.   Our formula is not identical to that found in Carlip \cite{Carlip}, however the energy-momentum tensor that he considered corresponds to the Kinnersley photon rocket \cite{Kinnersley:1969zz} which is slightly different from the energy-momentum used here. 

The first term in Eqn.\eqref{EffectiveGravitoField2} is the  special term that shows that the electric field for uniform motion is in the direction of the instantaneous position of the charge, something required by Lorentz invariance.   Of course the full gravitational field of a uniformly moving mass is not given only by Eqn.\eqref{EffectiveGravitoField2}, but will also contain gravitomagnetic type fields and the full set of fields can be obtained exactly by a simple Lorentz transformation of the Schwarzschild metric \cite{Aichelburg:1971xy}.

It should also be pointed out that Eqn.\eqref{EffectiveGravitoField11} is the calculation of the gravitational field of a single point mass.  A single point mass can actually only effect uniform, straightline motion as momentum must be conserved, and thus the RHS of Eqn.\eqref{EffectiveGravitoField11} can never be created except in the case Eqn.\eqref{EffectiveGravitoField2}.  If we want to consider more complicated motion, such as simple harmonic oscillations, we must add the field produced by a compensatory mass which is required by momentum conservation.  In the linear approximation, the fields simply superpose linearly.  We turn to the analysis of momentum conservation in the next section.

\subsection{Conservation of momentum in Newtonian mechanics\label{14}}
We have obtained the gravitational fields of a prescribed motion of a point source in Eqn.\eqref{EffectiveGravitoField11}.  However, if the motion is to be physical, as we have noted,  there has to a compensatory movement of a different source that maintains conservation of momentum.  Generally speaking the fields of the compensatory source will remove any dipole like gravitational fields that appear to have been created by the original source.  However the multipole expansion is made with respect to a fixed coordinate system common for both the system and the compensatory mass. 

We consider a system comprised of point sources whose center of mass is at position $\bm\xi$.  The compensatory mass or counterweight is also comprised of point sources whose center of mass is at position $\bm\zeta$.  These masses create gravitational fields which are detected/affect a point body located at position $\bm x_0$.  For  details, see Appendix A and especially  Fig.\eqref{test-mass-system-cw} which appears there.  The Newtonian definition of the center of mass is sufficient for our analysis as the radiation reaction terms are assumed to be completely negligible. However, it is clear that the positions of the masses could be such that it may not be possible to use the multipole expansion for the created gravitational fields as it may not be possible to satisfy simultaneously the required assumptions $|\bm\xi|\ll |\bm x_0|$ and $|\bm\zeta|\ll |\bm x_0|$ in any given coordinate system.  For example, the system may be close to the detector but the compensatory mass is by far the furthest away.  Then we would have  $|\bm\xi|\ll |\bm x_0|$ but $|\bm\zeta|\gg |\bm x_0|$, and for this case, the multipole expansion is not sensible, does not converge and cannot be made.  This will happen for example, if the centre of mass of the compensatory system is not close to the detector or if the compensatory mass is spread out over a relatively large region.  Such is the case when the oscillating mass is bolted to the earth, the compensatory mass being the somewhat large part of the earth  that reacts to the motion of the oscillating mass.  If we think of oscillations at time scales of a 10 Hertz to a $10^{4}$ Hertz \cite{Martynov2016},  the LIGO frequency band, the sound speed on the surface of the earth being in the range of a $\sim 500$  to $\sim 2000$ metres/second \cite{PhysRevD.58.122002}, the compensatory mass would be spread out over a volume with linear dimensions of a few metres (the distance to the floor from the detector to a $\sim 200$ metres.  The oscillating mass can be brought as close to the detector as possible, say $\sim 1$ cm, then the oscillatory mass would be close to the detector but the compensatory mass would be far, $|\bm\xi|\ll |\bm x_0|$ but $|\bm\zeta|\gg |\bm x_0|$ and the multipole expansion would fail.  This scenario has been analyzed in Thorne and Winstein \cite{Thorne:1998hq} for the case of humans walking near the detector,  and where the compensatory mass is treated as the local part of the floor and the earth below the building.  The reaction of the floor and earth is treated using elasticity theory.  As the compensatory mass is found at all different distances from the detector, and these distances are much greater than the distance between the detector and the human, one does not get the exact cancellation of the dipole terms as in Eqn.\eqref{mpe2} in Appendix A.

The conclusion to draw from this analysis is that for a physical, complete gravitating system (system and counter weight) made up of only positive masses,  the total dipole moment can at most be a linear function of time due to momentum conservation and  can be made exactly zero by choosing the appropriate, inertial, coordinate system.   However, the multipole expansion is dependent on the coordinate system, and the expansion only converges and is useful if the observation point $\bm x_0$ is the largest relevant distance in the expansion.  If this is not the case, then it can be that higher multipoles dominate, and the multipole expansion is not useful, even though it always remains true in any inertial coordinate system that the total dipole can at most be a linear function of time.  

For the sake of simplicity, let us imagine that the oscillating mass $m$ is attached to the compensatory mass $M$ by a long, thin, rigid rod, coupled with a spring, all in the horizontal direction.  Let the  rod be of length $N |\bm\xi|$ such that $\bm\zeta= N\bm\xi$, and making the assumption that the oscillating mass and the compensatory mass can both be taken as concentrated at their respective centres of mass,  the Newtonian potential at the detector is given by
\beq
\phi_{\rm Newton}(\bm x_0,\bm x^i,\bm y^A)\approx-\frac{Gm}{\left|\bm x_0 -\bm \xi\right|}-\frac{GM}{\left|\bm x_0-\bm \zeta\right|}=-\frac{Gm}{\left|\bm x_0 -\bm \xi\right|}-\frac{GM}{\left|\bm x_0-N\bm \xi\right|}.
\eeq
It is clear that the multipole expansion is not convergent and hence not useable for large enough $N$, but we can obviously see
\bea
\phi_{\rm Newton}(\bm x_0,\bm x^i,\bm y^A)&\approx& -\frac{Gm}{\left|\bm x_0 -\bm \xi\right|}+o\left(\frac{1}{N}\right)\nn
\ame -Gm \frac{1}{\left|\bm x_0\right|}  -\frac{G\bm x_0}{\left|\bm x_0\right|^3}\cdot\lb{m\bm \xi}\rb +o\left(\frac{1}{N}\right)+\cdots .
\eea
Neglecting the terms that are $o\left(\frac{1}{N}\right)$ we see that a dipole-like contribution of the motion of the oscillating mass does indeed give a non-vanishing contribution that in fact dominates over the quadrupole-like contribution.

\section{Experimental proposals}


\subsection{First proposal}

\begin{figure}[h!]
    \centering
    \includegraphics[scale = 0.5]{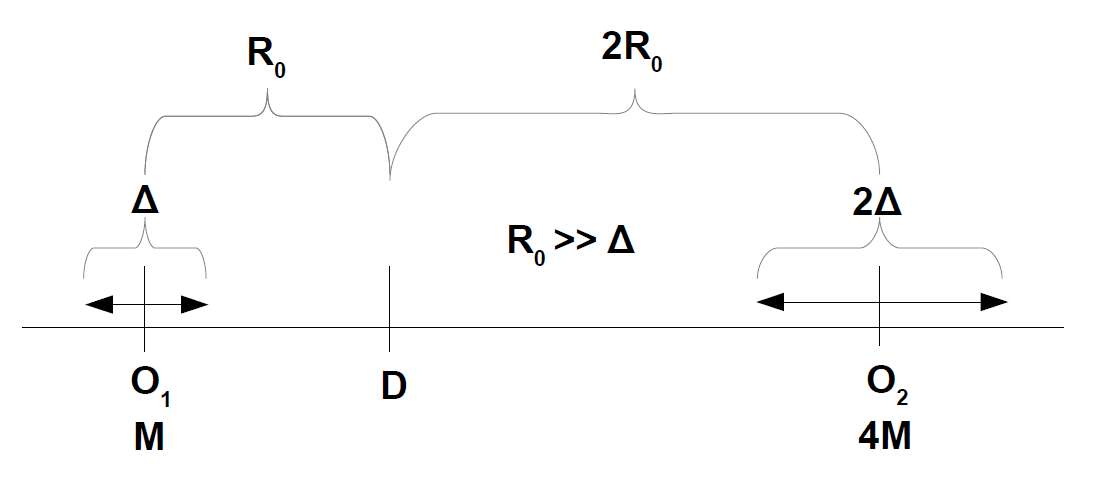}
    \caption{Schema of the experiment with two  oscillators ($O_1$, $O_2$) of respective mass ($M$, $4M$) and oscillating at distances ($R_0$, $2R_0$) with amplitudes ($\Delta$, $2\Delta$). }
    \label{fig:my_label}
\end{figure}

In this subsection we offer a (corrected) detailed analysis of the proposed system of  \cite{ParanjapeSpeedGravity}\footnote{In  \cite{ParanjapeSpeedGravity}, both the lack of aberration and the necessity of conservation of momentum were not taken into account.}.
The system analyzed in  \cite{ParanjapeSpeedGravity} corresponds to the following experimental configuration: a detector of gravitational phenomena (forces, waves, etc) has on each side (left and right), masses (which move) such that the Newtonian (instantaneous) gravitational forces produced at the detector exactly cancel (\ie the ratios of mass over distance squared to the detector for each side are chosen to be equal).  The simplest motion one can imagine is harmonic oscillation, with the left side mass $M$ at a distance $R_0+\Delta(t)$ and the right side mass $4M$ at distance $2R_0+2\Delta(t)$ (see Fig. \ref{fig:my_label}).   The amplitude of the oscillation is $\Delta$ on the left and $2\Delta$ on the right with $\Delta\ll R_0$.  The oscillation is synchronous so that's the Newtonian gravitational terms exactly cancel. The actual dynamical terms are in fact different and do not cancel, as we will see, and this affords the possibility of measuring the speed of propagation of gravitational effects.
The Newtonian theory is incomplete because it does not encompass any dynamical, time dependent effects of gravitation.  It posits instantaneous action at a distance, which is surely not accurate.  Any dynamical theory of gravitation, and we will be exclusively concerned  with Einsteinian general relativity, will suggest that gravitational effects have a finite propagation speed, and for any relativistically invariant theory, that propagation speed will be equal to the speed of light.  Consequently, any dynamical effects of gravitation will be perceived by an observer at what is termed the ``instantaneous time'' corresponding to the creation of those effects at the position of the source at what is termed the ``retarded time''.  The retarded time $t_r$ and the instantaneous time $t$, are defined implicitly by the equation
\begin{equation}
 t_r = t - \vert R(t_r) \vert/c_g, \label{tr}  
\end{equation}
where $R(t_r)$ is the position of the source at the retarded time, and is graphically represented on Fig. \eqref{Times}.
\subsubsection{Lack of aberration}
A naive analysis of the notion of retarded time would lead one to believe that to an observer, sources of gravitation always point back to their retarded position. For example, on this basis,  Laplace \cite{laplace1799traite} (in 1799 no less) concluded the speed of gravity must be greater than $10^{8}c$ based on the instability of the solar system as angular momentum was no longer conserved, or more recently van Flandern \cite{VanFlandern} computed from experimental data that the speed of gravity should be greater than $2 \times 10^{10} c$. This conundrum is known as aberration, or more precisely, the lack thereof for unaccelerated motion.  The claims in \cite{laplace1799traite} and \cite{VanFlandern} have been debunked, the actual physics is well understood and explained by Carlip\cite{Carlip} and Will \cite{Will:2003yj}.  Our analysis has to take into account this subtlety to compute the true effect of the moving masses at the detector. 
\subsubsection{Cancellation due to  momentum conservation}
A further subtlety arises due to momentum conservation.  The arbitrary motion of a given mass, say as in our proposal of simple harmonic motion, is simply unphysical.  There has to be a compensatory mass whose motion takes into account energy-momentum conservation.  The effect of such a compensatory mass, however large or complicated in its spread, is to remove the total dipole contribution to the gravitational forces.  On general grounds the total dipole contribution must be absent.  However, if the compensatory mass can be placed very far away from the detector, then its contribution can be taken as negligible.   The multipole expansion of both the oscillating mass and the compensatory mass does not make sense, and just the dipole contribution of the oscillating mass can contribute.  This is the import of the previous section on the conservation of momentum.

\subsubsection{Expression in terms of the instantaneous time}
Finally, a further complication arises from the fact that the two masses have different retarded times for a given instantaneous time, as their positions $R(t_r)$ and the corresponding retarded times $t_r$ are different for the two sides. Hence to combine their contributions at the detector, one must express each contribution in terms of the instantaneous time.  This requires inverting the expression for the retarded time, and any functions thereof,  in terms of the instantaneous time.    To do this inversion, we make use of the Lagrange inversion theorem \cite{Lagrange1770}.  The fundamental time dependent function that we must express in terms of the instantaneous time is the distance from the source to the observer/detector (placed at the origin of the coordinate system)
\beq
\bm{R}(t_r) = -\bm{r}(t_r) \equiv R_0 (1 + \alpha \sin (v t_r), 0, 0).
\eeq

\begin{figure}[h!]
\begin{center}
\begin{tikzpicture}[line cap=round,line join=round,>=triangle 45,x=1cm,y=1cm]
\clip(0,0) rectangle (6,6.5);
\draw[line width=1pt,smooth,samples=100,domain= -1 :10, rotate=90] plot(\x,{0.5*sin((5*(\x))*180/pi)-1});
\draw [line width=1pt] (1,-1)-- (1,7); 
\draw [line width=1pt] (4,-1)-- (4,7);

\draw [line width=1pt] (1.5,2.2)-- (4,2.2);  
\draw[color=black] (2.5,1.9) node {$R(t_r)$};
\draw[color=black] (4, 2.2) node[right] {$t_r$};

\draw [line width=1pt] (1.3, 5.77)-- (4,5.77);
\draw[color=black] (2.5, 5.5) node {$R(t)$};
\draw[color=black] (4,5.77) node[right] {$t$};

 \draw [line width=1pt] (1.5,2.2)-- (4,5.77);
\draw (2,2.2) arc (0:53:0.5);  
\draw[color=black] (2.3,2.6) node {$\dfrac{\pi}{4}$};
 
\draw [->, line width=1pt] (2.5,0.5)-- (4,0.5);
\draw [->, line width=1pt] (2.5,0.5)-- (1,0.5);
\draw[color=black] (2.5,0.75) node {$R_{0}$};

\end{tikzpicture}
\caption{Sketch of the different times and distances and the links between them for an oscillating motion.}
\label{Times}
\end{center}
\end{figure}

We define the  dimensionless retarded time, $z$, as
\begin{equation}
    z := \frac{c_g t_r}{R_0},
\end{equation}
where $R_0$ is a fiducial distance, taken as in Fig.\eqref{fig:my_label}, the distance from the left side mass to the detector at $t_r=0$.  We imagine a harmonic motion as $R(t_r)=R_0+\Delta\sin(\omega t_r)$ where we impose that $\Delta \ll R_0$, then $\al = \Delta/R_0 \ll 1$ can act as the expansion parameter. Then the motion is given by 
\begin{equation}
    R(z) = R_0 (1 + \alpha f(z)),
\end{equation}
where $f(z)$ describes the oscillation around $R_0$. The retarded time equation  $t_r = t - \vert R(t_r) \vert/c_g$ defining $y = \frac{c_gt}{R_0} - 1$ becomes
\begin{equation}
    z = y - \alpha f(z) \label{44}
\end{equation}
The minus 1 in the definition of $y$  just corresponds to the light travel time for the distance $R_0$ and correspondingly,  $\al$ becomes the expansion parameter.  Eqn.\eqref{44}  is seen as an equation implicitly defining the retarded time $z$ in terms of the instantaneous time $y$.  The inversion of Eqn.\eqref{44} for $z$, or in fact any function $g(z)$ in terms of $y$ is given in a series expansion by the Lagrange inversion theorem.

\subsubsection{Lagrange inversion theorem}
To combine the effects of the motion of different sources located at different positions at different retarded times, we must express their effects in terms of the instantaneous time.   The Lagrange inversion theorem is a perfectly suited formula for this goal and gives an expansion of functions of the retarded time in terms of the instantaneous time. Although we will only use the first few terms of the expansion in this article, we think it is useful to the reader to know the full expansion.  The proof of the theorem and all details can be found, for example, in Whittaker and Watson \cite{Whittaker}.
\begin{theorem}
Let $f(z)$ be a function of $z$ which is analytic on and inside a contour $\mathbf{C}$ surrounding a point $y$, and let $\alpha$ be such that the inequality
\begin{equation*}
    \vert \alpha f(z) \vert < \vert z - y \vert
\end{equation*}
\textit{is satisfied at all points $z$ on the perimeter of $\mathbf{C}$; then the equation}
\begin{equation}
    z = y + \alpha f(z)
\end{equation}
\textit{regarded as an equation in $z$ has one root in the interior of $\mathbf{C}$; and further any function $g(z)$ analytic on and inside $\mathbf{C}$ can be expanded in a power series in $\al$ by the formula}
\begin{equation}
    g(z) = g(y) + \sum_{n=1}^{\infty} \frac{\alpha^n}{n!} \frac{d^{n-1}}{dy^{n-1}} \left( g'(y) f^n(y) \right).
    \label{LagrangeInversionThm}
\end{equation}
\end{theorem}
This is the general theorem, we will use this expansion only to second order in this article. 
\subsubsection{Application to the first proposal}
We will apply the theorem to the motion and effects of  the mass on the left, $O_1$.  We have the retarded position
\beq
\bm{R}(z)  \equiv R_0 (1 + \alpha \sin (v z), 0, 0),
\eeq
where $v = \frac{\omega R_0}{c_g} \ll 1$ is the dimensionless speed and evidently $v z = \frac{\omega R_0}{c_g} \cdot \frac{c_g t_r}{R_0} = \omega t_r$ and the acceleration produced from Eqn.\eqref{EffectiveGravitoField2} is given 
\begin{align}
\frac{d\bm p}{dt}&=  \frac{-Gm M \bm{\hat{x}}}{R_0^2} \left[ \frac{ 1 + \alpha v \cos(vz)  -3 \alpha v^2 \sin(vz)  - 3 \alpha^2 v^2 \sin^2(vz)  - 3 \alpha^2 v^2 \cos^2(vz)    }{(1+\alpha \sin(vz)  )^2 (1 + \alpha v \cos(vz)  )^3 \sqrt{1 - \alpha^2 v^2 \cos^2(vz) }}\right].\label{48}
\end{align}
The gravitational force is all expressed in the retarded time here, the application of the Lagrange inversion theorem is then quite straightforward but tedious.  The detailed computation is found in Appendix B, however, the general idea is clear.  We apply the  theorem to the RHS of Eqn.\eqref{48}, and re-express it in terms of $y$ the instantaneous time.  Throughout the computation we  Taylor expand in $v$ and $\al$.   Subsequently we write the trigonometric functions of $v y$ in terms of $\omega t$ note that there is a shift by -1 given in Eqn.\eqref{44} in the definition of the instantaneous time and $y$.   Finally, Eqn.\eqref{48}  as a function of the instantaneous time $t$, is found to be
\begin{align}
   \frac{d\bm p}{dt}(t)&=  -Gm M \bm{\hat{x}} \left[ \frac{1}{R(t)^2} - \frac{1}{R_0^2} (4\al v^2 \sin(\omega t) + \frac{5}{2}\al^2 v^2 \cos^2(\omega t) - 4\al^2 v^2 \sin^2(\omega t) ) \right] \nonumber \\
    &=  -Gm M \bm{\hat{x}} \left[ \frac{1}{R(t)^2} - \frac{1}{R_0^2} \left(4a(t) + \frac{5}{2}\beta(t)^2 - 4 a(t) \Delta(t) \right) \right]\label{egff}
\end{align}
where the velocity is given by $\beta(t)=\al v\cos(\omega t)$ and the acceleration is given by $a(t) = \al v^2 \sin(\omega t)$.  The expression in terms of $\beta(t)$, $a(t)$ and $\Delta(t)$ is only valid for the harmonic motion that is considered.   

The first term is the Newtonian instantaneous term arriving because of Lorentz invariance in the absence of acceleration and  relativistic corrections.  The next term is proportional to the acceleration $a(t) = \al v^2 \sin(\omega t)$. This term comes from the dipole moment of the oscillating mass.   However, the motion of a single, simple harmonic oscillator does not conserve momentum and simply does not occur physically.  This is solved by adding a compensating mass moving synchronously  in the opposite direction. For example, if $O_1$ is bolted to the floor, the compensatory mass is effectively the Earth.  For the sake of simplicity and clarity, consider just a very heavy mass, $O_H$, moving in the opposite direction to $O_1$, then $M \dot{\Delta}(t) + M_H \dot{\Delta}_H(t) = 0$ for momentum conservation.  Taking the compensatory mass at a distance $R_H$ with vibrational amplitude $R_H\alpha_H$, we get
\beq
MR_0\alpha -M_HR_H\alpha_H=0
\eeq
giving
\beq
\alpha_H=\frac{MR_0}{M_HR_H}\alpha.
\eeq
Then for the gravitational effect of the heavy mass, we can simply use the same formula as Eqn.\eqref{egff} replacing $\al\to\alpha_H = -\frac{MR_0}{M_HR_H} \alpha$  and $v\to v_H=\frac{\omega R_H}{c_g}$ we find that the dipole term of the compensating mass is given by
\beq
\frac{GM_H}{R_H^2}\bm \hat x 4\alpha_H (v_H)^2\sin(\omega t)=\lb\frac{R_0}{R_H}\rb\frac{GM\bm \hat x 4\alpha v^2}{R_0^2}\sin(\omega t)=\lb\frac{R_0}{R_H}\rb \frac{GM\bm \hat x }{R_0^2}4a(t).
\eeq
 Hence for $\lb\frac{R_0}{R_H}\rb$ taken small enough, we can drop the dipole contribution of the compensating mass.  Thus only the dipole term of the oscillating mass gives the dominant contribution.  We would like to stress that this in no way means that the total dipole moment is contributing to the oscillating gravitational field at the detector.  Its contribution, because of momentum conservation, has to trivial in any inertial coordinate system.   The contribution of the subsequent terms in hte multipole expansion due to the compensatory mass are smaller by an additional factors of $\alpha_H$ and hence are in principle utterly negligible.  Therefore the only contribution of the compensatory mass that we will keep is its instantaneous Newtonian monopole term.  Thus we find for the acceleration of the detector due to the masses on the left is given by 
\begin{align}
       \frac{d\bm p}{dt}&= -m\bm{\hat{x}} \left[ \frac{G M_H}{R_H(t)^2} +\frac{G M}{R(t)^2} - \frac{G M}{R_0^2} \left( 4a(t)+\frac{5}{2}\beta(t)^2 - 4a(t)\Delta(t) \right) \right]
\end{align}
This is the net effect of the left-side oscillating masses ($M$ and $M_H$) on a detector placed at the distances $R_0$ and $R_H$ respectively from the detector.

The system on the other side of the detector is composed of mass $4M$ at a distance $2R_0$ with oscillation amplitude $2\Delta$ and of course a compensatory mass $4M_H$ placed at $2R_H$. The choice of the right side masses is made in order to cancel the instantaneous Newtonian force of all the masses on the detector. The computation of the right side system is identical to the left side system since the dimensionless parameters all have the same values and only the direction of the forces created are in the opposite direction and the values of the dimensionless velocities are doubled.  Then total acceleration of the detector is given by

\begin{align}
  \frac{d\bm p}{dt}(t)&= -m\bm{\hat{x}}\,\, 4 \left( \frac{GM }{R_0^2} \right) \left( 4a(t)+\frac{5}{2}\beta(t)^2 - 4a(t)\Delta(t) \right) +m\bm{\hat{x}}   \left( \frac{G M}{R_0^2} \right) \left(4a(t)+\frac{5}{2}\beta(t)^2 -4 a(t)\Delta(t) \right) \nonumber \\
&=  -m\bm{\hat{x}} \frac{ G M}{R_0^2}  \left( 12a(t)+ \frac{15}{2} \beta(t)^2 - 12 a(t)\Delta(t) \right).\label{txef}
\end{align}

We note that that the result we have obtained is quite different from that obtained for the field in the wave zone, which corresponds to distances much larger than the size of the source and the wavelength of the radiation produced.  In the wave zone the metric perturbations drop off like $\sim \frac{1}{r}$ while the corresponding gravitational fields then fall off as $\sim \frac{1}{r^2}$.  This behaviour is simply not valid in our case.  Our result is not proportional to the third time derivative of the quadrupole moment, which is the result obtained as the leading term in the wave zone.  Here we are well inside a single wavelength, we are computing what is normally called Newtonian noise. 

\subsubsection{Measurement}
The effect of the oscillating gravitation field on a detector will be to force oscillations, simply according to Newton's law.   We have
\beq
\frac{d\bm p}{dt}=m\frac{d^2\bm X}{dt^2}=-m\bm{\hat{x}} \frac{ G M}{R_0^2}  \left( 12a(t)+  \frac{15}{2} \beta(t)^2  -12 a(t)\Delta(t)\right).\label{acc}
\eeq
As expected the mass of the detector $m$ cancels from this equation as dictated by the Principle of Equivalence.  Then the spatial motion of the detector is obtained by integrating Eqn.\eqref{acc} twice.   We note that $a\sim 1/c_g^2$ and $\beta\sim 1/c_g$ all terms in Eqn.\eqref{acc} contain factors of $1/c_g$.  Hence if there was truly no aberration and the propagation speed was infinite, \ie $c_g\to\infty$,  our result would of course vanish.   Hence the actual measurement of any effect would confirm a finite propagation speed for gravity, while precision measurements could be used to determine $c_g$.  Our calculation is the first for the dynamical effects of gravitation at laboratory sized distances, the instantaneous effects have been designed to exactly cancel.  Previous calculations have always neglected the finite propagation speed of gravitational effects at these distances.  As we will see, the dynamical effects are not utterly negligible.

\subsubsection{Optimization}

The motion we have studied allows for the choice of a number of parameters, $\Delta,\omega,M,R_0,$ and $\rho, r_s$ the density and size of the oscillating masses.   This gives a window for optimizing the resulting force on the detector.  However, there are different constraints that must be taken into account.  We will not analyze the compensatory masses, they must exist, but their net effect on the detector are taken to be vanishing or negligible.

The first constraint comes from imposing the cancellation of the Newtonian terms (and has already been imposed in obtaining Eqn.\eqref{txef}) and leads to 
\begin{equation}
    \frac{M}{R_0^2} = \frac{M'}{{R'} _0^{2}}, \ \ \ \ \ \frac{\Delta}{R_0} = \alpha = \frac{\Delta '}{R'_0} \ \ \ \ \text{and} \ \ \ \ \omega = \omega'
\end{equation}
where the unprimed and primed quantities are specific to the systems on either side of the detector.  We also impose that $\al\ll 1$ , especially since we neglect terms of order $\al^3$ in the computation\footnote{It is easy to take into account higher orders terms in $\al$.}.

The second constraint is due to physical considerations. The masses, which were taken as point masses during the computation, should actually be spherical bodies made of the some material with density $\rho$. Therefore physically, for a spherical mass of density $\rho$ its radius $r_s$ must be smaller than the distance to the detector, $R_0$ : 
\begin{equation*}
    r_s = \zeta R_0 \ \ \ \text{with} \ \ \ \zeta < 1.
\end{equation*}
The $\zeta$'s of the two masses are different but linked by the equality $ \zeta'^{3} = \zeta^3\cdot R_0/R'_0=\zeta^3 \cdot \Delta /\Delta'$ from imposing the cancellation of the instantaneous Newton force. 

Then our expression for the force becomes
\bea
\frac{d\bm p}{dt}(t)&=&   -mG \frac{4\pi\rho R_0^3\zeta^3}{3} \bm{\hat{x}} \frac{ \al \omega^2 }{c_g^2}\lb 12\sin(\omega t) +\alpha\lb \frac{15}{2} \cos^2 (\omega t)-12 \sin^2(\omega t) \rb\rb\nn
  \ame   -mG \frac{4\pi\rho R_0^3\zeta^3}{3}\frac{ \al \omega^2 }{c_g^2} \bm{\hat{x}} \lb 12\sin(\omega t)+\al \lb\lb6+ \frac{15}{4} )\cos (2\omega t)\rb-\lb 6- \frac{15}{4} \rb\rb\rb.
\eea
Dropping the terms proportional to $\al^2$, we have an oscillating driving force on the mirror that is a consequence of the relative delay between the signal propagating from the sources to the mirror.  The Newton equation for the motion of the mirror is simply
\beq
\frac{d^2X}{dt^2}=G \frac{4\pi\rho\zeta^3R_0^3 \al \omega^2 }{3c_g^2}   12\sin(\omega t) 
\eeq
which integrates trivially by taking out the factor of $\omega^2$ in the numerator
\beq
X(t)= -G \frac{4\pi\rho\zeta^3R_0^3 \al  }{3c_g^2} \sin(\omega t).
\eeq
Interestingly, the frequency plays no role in the observability of the effect. 
Writing $M_0= \frac{4\pi \rho R_0^3}{3}$ we have
\beq
X(t)= -G \frac{\zeta^3M_0 \al  }{c_g^2}  \sin(\omega t) .
\eeq
Taking nominally $\zeta=.9$, $\al=.1$, we find the overall numerical factor gives
\beq
X(t)= -0.72 G \frac{M_0  }{c_g^2}  \sin(\omega t).
\eeq
Taking $c_g=c=2.99\times 10^{8}$ and $G=6.67\times 10^{-11}$ in MKS units we get in metres
\beq
X(t)= -0.72\times 7.46\times 10^{-28} M_0 \sin(\omega t).=5.05\times10^{-28}M_0 \sin(\omega t).
\eeq
For a large mass of say 10 metric tons, $M_0=10^{4}kg$, the amplitude of the oscillations of the mirror are $X(t)\sim 10^{-24}m$.  This is outside of the range of present technology which allows the measurement of amplitudes of $\sim 10^{-21}m$.  However we do not see our predicted amplitude as totally out of the realm of possibility, in the hopefully not too distant future. It should be noted that  recent analyses \cite{Parikh:2020nrd,parikh2020signatures} of the possibility of observing gravitational quantum fluctuations, that would establish unequivocally the existence of quantum gravitons, require observations at the level of $10^{-35}m$, which is substantially smaller than our projection.  We speculate that the handful of  orders of magnitude improvement in measurement capability could be surpassed in  future generation gravitational wave detectors. 

\subsection{Second proposal}
Here we explore the possibility that the acceleration of the source system could give rise to the dominant effect.  If we neglect all velocity dependent terms, we are left with, from Eqn. \eqref{EffectiveGravitoField11}, dropping all terms that cancel with the mass on the other side and the compensatory masses
\beq
\frac{d\bm p}{dt}=
     -mG M \left[\frac{\bm{R}}{ R^3} (  \bm{R}\cdot \dot{\bm{\beta}}) -\frac{4\bm{\dot\beta}}{R} \right]_{ret}=  -mG M \left[\frac{\bm{n}}{ R} (  \bm{n}\cdot \frac{d^2{\bm{r}}}{dt^2}) -\frac{4}{R}  \frac{d^2{\bm{r}}}{dt^2} \right]_{ret}.
   \label{EffectiveGravitoField12}
\eeq
Then the geodesic equation for the detector (mirror in LIGO), dropping the vectorial notation as we take everything to move in one direction and the $\kappa$ factor, will be
\beq
m\frac{d^2X}{dt^2}=mG M \left[\frac{3}{ R}  \frac{d^2{{x}}}{dt^2}  \right]_{ret}.
\eeq
If we imagine that the system suffers an impulse, a large acceleration for a short period of time followed by a period of coasting, such as a steel ball bearing bouncing between two fixed walls with a relatively large and essentially constant retardation time, for example, then we can estimate the effect on the detector by integrating this equation  with the retarded time put in the right hand side.  As both sides are identical in time derivatives, we get
\beq
X(t)=3G M \left[\frac{x(t)}{R}   \right]_{ret}.\label{acce}
\eeq
Noting that there is a $1/c_g^2$ implicit in the right hand side, we find in metres
\beq
\left| X\right|\sim 7.46\times10^{-28}M \alpha 
\eeq
where $\alpha$ is the fractional amplitude of the accelerated motion.  
This expression is of course only the contribution of one side, and at the corresponding retarded time, however, it is not cancelled by the other side due to the retardation effects.  It is still rather small, and with $\al=.1$, and a mass of $10^{4}kg$, which is very large but not absolutely impossible, we have
\beq
\left| X\right|\sim 7.46\times10^{-25}m.
\eeq
This amount of  disturbance is again, not yet measurable, however, it is also not completely out of the realm of possibilities in the future. 
\subsubsection{Gravitational earthquake detection}
An even more serendipitous potential observation has to do with earthquakes.  Gravitational observation of earthquakes has been discussed in the literature \cite{HarmsJ2015Tgpi,cdi_oup_primary_10_1093_gji_ggaa486,JuhelK2018EEWU,ww}.  However the theoretical analysis done in these references do not consider the dynamical aspects of the nature of gravitation.  All calculations are done in the assumption that the speed of gravitational effects is essentially infinite.  
x
For large earthquakes the analysis of small seismic disturbances done in \cite{PhysRevD.58.122002} is not relevant.  In magnitude 8 and higher earthquakes, a huge mass, part of a tectonic plate can move of the order of several tens of meters in a short period of time, tens of seconds.  Although the accelerations experienced are not large,  the mass that moves can be so large that the gravitational effect is potentially observable.  Indeed it may well be possible to detect large magnitude earthquakes gravitationally, well before before their seismic signal arrives.  

During a magnitude 8 or higher earthquake,  part of a tectonic plate which normally has a thickness of around $100$km can move (locally) a distance of the order of 40 metres \cite{Bletery1027,SCHELLART201341}.  The thickness of the plate can vary, however, around 100km is a reasonable estimate.  The tectonic plates form a large jigsaw puzzle that covers the surface of the earth, and any motion of the plates is highly constrained. When an earthquake happens, most of a plate does not  move, only built up stresses in a local region are released and only the part of the plate in the local region of the earthquake actually moves.  The motion of mass corresponds to  the liberation of strains and stresses that have been slowly built up  over many years.  The movement can be quite dramatic.  The subduction (one plate moving under a neighbouring plate) of the Pacific plate during the recent (April, 2011) earthquake in Fukushima, Japan corresponded to a local displacement of about $40$ metres.  Did the whole Pacific plate move  by 40 metres?  Of course not, only a small, local portion of the plate and the corresponding mass, moved this distance, that region which is called the subduction zone for this type of earthquake.  

For a magnitude 8, an earthquake can last several tens of seconds and mass motion occurs along a distance of about 100km.  For a magnitude 9 or higher earthquake, stresses can be released over a 1000km distance and the quake can last for up to 5 minutes \cite{nz}.   For the example of the Fukushima earthquake, which was the second largest ever recorded (after the 1960 Chile earthquake) the mass in the subduction zone of the Pacific plate moved of the order of 40 metres under the Eurasian plate.  

Let us estimate the volume of the part of the plate that moved in the Fukushima earthquake, as having length 100km, a thickness of 100km and a width of 40 metres, which corresponds to  $4\times10^{11}$ cubic metres.  The density of the lithosphere (the solid crust of the earth, that makes up the tectonic plates) is approximately $2800 $kg per cubic metre \cite{bc}, therefore we get a mass of $1.12\times 10^{15}$kg.  

We do have to think whether conservation of momentum would simply, exactly cancel the existence of any such effect coming from the movement of this mass.  Evidently, such a motion cannot occur alone, while respecting conservation of momentum.  The reaction of the rest of the Earth to the movement of this mass due to the release of the built up stress, reverberated throughout the Pacific plate, this is why we actually feel earthquakes.  The entire Pacific plate and probably various other parts of the Earth reverberated so as to respect conservation of total  momentum.  However, we have understood that the motion of the distributed parts reacting due to conservation of momentum of a given specific motion, do not act in an identical fashion to the specific motion,  and hence do not exactly cancel the gravitational effects on a detector because the distance to the detector is not the same.    Additionally the magma under the Eurasian plate which allowed for the intrusion of the Pacific plate, had to flow around the inserted plate, to make available the volume of the inserted plate. The motion of this mass again would be distributed around the inserted plate, and would create gravitational perturbations at the detector.  However, as noted before, these contributions would be distributed over a large regions of the earth compared to the location and size of the initial intrusion, and the resulting gravitational effects would never exactly cancel the original effect. 

Let us calculate the effect of such a motion on a mirror as in the Ligo detector.  
From Eqn.\eqref{acce}
\beq
X(t)=-\frac{3G}{c_g^2} M \left[\frac{x(t)}{R}   \right]_{ret}\sim \frac{3\times 6.67\times 10^{-11}\times 1.12\times 10^{15}}{(2.99\times 10^{8})^2}\left[\frac{x(t)}{R}   \right]_{ret}=2.51\times 10^{-11}\left[\frac{x(t)}{R}  \right]_{ret}\label{seismic}
\eeq
Taking $\left|\left[\frac{x(t)}{R}  \right]_{ret}\right|\sim 40/10^{6}$ for a magnitude 9 or greater earthquake with a movement of $40m$ and occurring at $R\sim 1000$km away, we have
\beq
\left| X\right|\sim 1.00\times10^{-15}m.
\eeq
which is very, easily observable.  Of course the estimated mass and other parameters could vary considerably, however, we can afford a diminution of our estimate by several orders of magnitude, but the effect is still observable.  We have also not taken into account the effect of the compensatory mass, which is this case would be the rest of the earth.  But applying, the rule of thumb that can be obtained from the calculations done in \cite{Thorne:1998hq} using elasticity theory for the reaction of the rest of the earth to humans walking, to the motion due to the earthquake, we can imagine that the cancelling effect of the movement of the rest of the earth will only be an effect of the same order of magnitude.   This means that  we would expect that our calculated gravitational effect would be modified by terms that would be of the same order of magnitude, but which would not exactly cancel the calculated effect.  Therefore the effect calculated in Eqn.\eqref{seismic} would only be affected by terms of the same order of magnitude, changing the amplitude by at most a factor of order 1.  

It would seem that the LIGO type detector is an ideal early warning sensor for large earthquakes, gravitational effects presumably propagate at the speed of light which is much faster than the speed of seismic waves.  Having many such detectors situated around the globe, as is the actual case, would allow for quick referencing data on the actual position of the earthquake.  

\section{Conclusion}
We have proposed an experiment that could observe the dynamical effects of general relativity in the near zone and could be used to measure the speed of gravity in a directly controllable laboratory setting.  It would be important to be able to do this measurement since up to now, the measurement of the speed of gravitational propagation is solely based on production at astronomically distant sources and the subsequent indirect measurement of the speed by comparison with the arrival time of electromagnetic radiation, in principle produced simultaneously by the same source.  Such an indirect measurement is surely perfectly good, especially as it confirms that the speed of gravity and the speed of light are equal to one part in $10^{15}$.  However, an incontrovertible measurement would correspond to the production of the propagating gravitational disturbance in the laboratory and the measurement of the elapsed time before its subsequent arrival at a detector also in the laboratory.  The experimental proposals that we have analyzed are not possible at the present, however, they are neither beyond the not so far horizon of future possibilities.

In the process of our calculation, we analyzed the role of instantaneous and retarded time and focused on how to pass from one to the other.  We use a mathematical result, the Lagrange inversion theorem, that could be  useful in future computation. It could also be easily adapted for use even in pure electromagnetism situation.

Two subtleties have been addressed which are at the heart of the reason why the dynamical effects of gravitation are so difficult to observe, apart from the obvious fact of the intrinsic weakness of gravity due to the very small coupling constant. First the fact that there is no aberration in the gravitational fields for uniformly moving sources (masses).  This is a subtle consequence of Lorentz invariance.  Therefore, only accelerated motion can give rise to dynamical retardation effects of the gravitational fields, and consequently the observable effects are extremely small.  Second is the fact that energy-momentum conservation denies the possibility of a dipolar dynamical field.  The gravitational dipole must be time independent and actually origin dependent.  Therefore the first non-zero dynamical effects can only be  observed  in the quadrupole gravitational fields, which results in a further diminution of  potentially observable effects.  However, the usual higher power decay of the quadrupole for the far zone, is not valid in the near zone, hence there is some hope that the effects in the near zone, are  not suppressed into oblivion.  Although too weak at the present time, nevertheless, we still hope that  the dynamical effects of gravity  could be observed in special purpose, extremely high precision detectors of  gravitational acceleration, in the not too distant future.

Finally, we have given an analysis of the signal measured due to the acceleration of different masses.  Controlled masses in a laboratory setting again do not give signals that are presently measurable.  However, the enormous mass movement that occurs in some earthquakes would seem to give signals that are observable in interferometric gravitational wave detectors.  This fact has already been noticed  in the extant literature \cite{HarmsJ2015Tgpi,cdi_oup_primary_10_1093_gji_ggaa486,JuhelK2018EEWU,ww}.  Indeed, the ensemble of such interferometric gravitational wave detectors could serve as an early warning system for large earthquakes, with location information available by triangulation and signal arrival time, if there are sufficiently many such detectors.  
\section{Acknowledgements}
We thank NSERC of Canada  and the FacultŽ des Žtudes supŽrieures et postdoctorales and the DŽpartement de physique of the UniversitŽ de MontrŽal for financial support.  We thank Steven Carlip, Richard MacKenzie and Kip Thorne for useful discussions and correspondence.  This work received honourable mention in the 2012 Gravity Research Fondation Essay Competition where it was first enunciated.  
\section{Appendix A: Analysis of the conservation of momentum. \label{A}}
We will first consider the case of a set of particles (atoms, treated as point particles) with masses and positions $\{m_i, {\bm x}^i\}$ which correspond to our system that is creating the Newtonian noise.  These masses are compensated by a counter weight which is comprised also of a set of particles (also atoms, treated as point particles) with masses and positions $\{m_A, {\bm y}^A\}$.  ${\bm x}^i$ and ${\bm y}^A$ are the coordinates of the mass points with respect to a fixed origin.  The system could correspond to a human walking or a mass oscillating or any system of gravitational sources creating a desired, time dependent gravitational field.  However the required motion can require compensating movement of masses in order that momentum is conserved.  For example, an oscillating pendulum could be in a housing that is mounted on the ground.  In that the case, the counter weight would correspond to all the particles that make up the ground in a reasonable sized region around the spot where the housing is mounted.  The motion of a pendulum alone does not conserve momentum, and indeed, the housing and through the mounting, the particles in the ground supply the required momentum for conservation.   The  counter weight could also be  a specific macroscopic mass attached to the system, designed in such a way (usually, a very large mass) so that its motion will allow for conservation of momentum, however, its dynamical gravitational fields would be negligible.  We will see how this is possible in our example.  
\begin{figure}[!h]
    \centering
    \includegraphics[scale = 0.5]{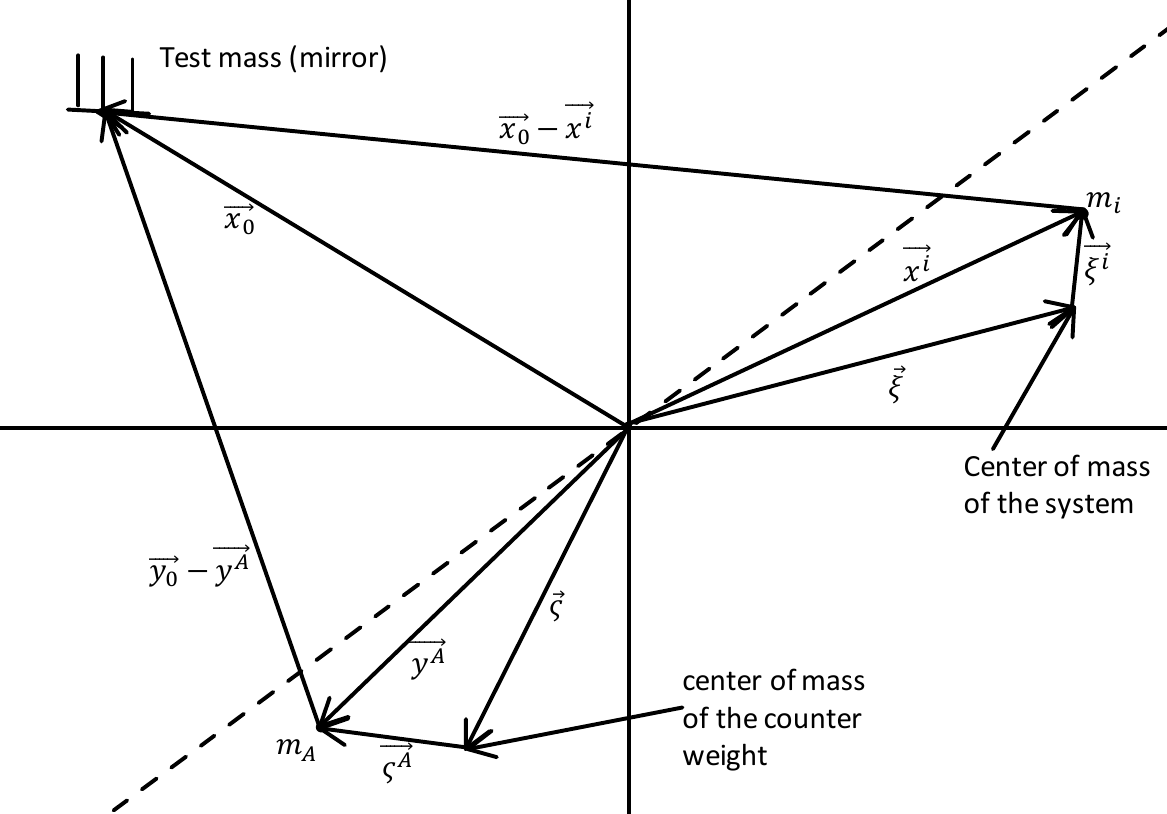}
    \caption{The scheme of the test mass (mirror), the system and the counter weight.}
    \label{test-mass-system-cw}
\end{figure}

The position of the centre of mass of the entire system and counter weight $\bm X$ is given by
\beq
\bm X =\frac{\sum_im_i {\bm x}^i +\sum_A m_A{\bm y}^A }{\sum_im_i +\sum_A m_A}.
\eeq
Let the mass of the system be $m=\sum_im_i$ and the mass of the counter weight $M=\sum_Am_A$.  We define $\bm \xi$ and $\bm \zeta$ to be the centre of mass of the system and the counter weight respectively
\beq
\bm \xi=\frac{\sum_i m_i {\bm x}^i}{m}\quad\quad\quad \bm \zeta=\frac{\sum_A m_ a{\bm y}^A}{M}.
\eeq
These should be identified as proportional to the dipole moment of the system and of the counter weight respectively.  The deviation of the particles making up the system and the counter weight from their respective centre of mass is defined as  ${\bm x}^i=\bm \xi +{\bm\xi}^i$ and ${\bm y}^A=\bm \zeta +{\bm\zeta}^A$.  Then it is easy to find
\beq
\bm X=\frac{m\bm \xi+M\bm \zeta}{m+M}+\frac{\sum_im_i{\bm\xi}^i+\sum_A m_A{\bm\zeta}^A}{m+M}.
\eeq
Now the dipole moment of the system is given by
\beq
\sum_im_i{\bm x}^i=\sum_im_i \bm \xi +\sum_im_i{\bm\xi}^i=m \bm \xi +\sum_im_i{\bm\xi}^i=m\frac{\sum_i m_i {\bm x}^i}{m} + \sum_im_i{\bm\xi}^i=\sum_i m_i {\bm x}^i +\sum_im_i{\bm\xi}^i.
\eeq
therefore
\beq
\sum_im_i{\bm\xi}^i=0
\eeq
and correspondingly 
\beq
\sum_Am_A{\bm\zeta}^A=0
\eeq
\ie the contributions of the deviations of the particles of the system from its centre of mass does not contribute to the dipole moment of the system, and correspondingly for the counter weight.
Then clearly
\beq
\bm X=\frac{m\bm \xi+M\bm \zeta}{m+M}.
\eeq
Total momentum of the system and the counter weight together must be conserved, therefore
\beq
\dot P=\frac{d}{dt}\lb\sum_im_i\dot { \bm x}^i+\sum_Am_A\dot{\bm y}^A\rb= \frac{d^2}{dt^2}\lb\sum_im_i {{  \bm x^i}}+\sum_Am_A {{  \bm y^A}}\rb=m\Ddot{\bm \xi}+M\Ddot{\bm \zeta}=0
\eeq
\ie
\beq
\Ddot{ \bm{X}}=0.
\eeq
Thus we can write
\beq
{ \bm{X}}=\bm{V}  t + \bm{X_0}
\eeq
and in the centre of mass system with the appropriate choice of origin, $ \bm{V}=0$ $ \bm{X_0}=0$, so that indeed we can take
\beq
{\bm{X}}=0.
\eeq
We assume that the deviations ${\bm\xi}^i$ and ${\bm\zeta}^A$ are small when compared to the respective centres of mass $\bm \xi$ and $\bm \zeta$ respectively.  Then any function of the coordinates can be expanded in a Taylor series about the undeviated coordinates.  We will specialize to functions that can be written as a linear superposition of a contribution coming from each mass point, as it is the kind of function that is relevant:
\beq
f(\bm x^i,\bm y^A)=\sum_i f_i(\bm x^i)+\sum_A f_A(\bm y^A).
\eeq
Then writing the coordinates of $\bm x^i$ as $x^i_\alpha$ where the greek index takes the values $\alpha=1,2,3$ and summation over repeated greek indices is assumed, etc. we have
\bea
f(x^i_\alpha,y^A_\beta)\ame f( \xi_\alpha+{\xi}^i_\alpha, \zeta_\beta+{\zeta}^A_\beta)=f( \xi_\alpha, \zeta_\beta)+\sum_i\partial_{\xi_\gamma} f_i( \xi_\alpha){\xi}^i_\gamma+\sum_A\partial_{\zeta_\gamma} f_A( \zeta_\beta) {\zeta}^A_\gamma+\nn
&+&\haf 1\lb\sum_i\partial_{\xi_\gamma}\partial_{\xi_\epsilon} f_i( \xi_\alpha){\xi}^i_\gamma{\xi}^i_\epsilon +\sum_A\partial_{\zeta_\gamma}\partial_{\zeta_\epsilon} f_A(\zeta_\beta){\zeta}^A_\gamma{\zeta}^A_\epsilon\rb +\cdots . 
\eea
Applying this to the Newtonian potential at a position $\bm x_0$, the position of a detector for example, we have
\beq
\phi_{\rm Newton}(\bm x_0,\bm x^i,\bm y^A)=-\sum_i\frac{Gm_i}{\left|\bm x_0-\bm x^i\right|}-\sum_A\frac{Gm_A}{\left|\bm x_0-\bm y^A\right|}.
\eeq
Thus $f_i(\bm x_0,\bm \xi)=-\frac{Gm_i}{\left|\bm x_0-\bm \xi\right|}$ and then  we have $\partial_{\xi_\gamma}f_i(\bm x_0,\bm \xi)=-Gm_i\frac{ {x_0}_\gamma-\xi_\gamma}{\left|\bm x_0-\bm \xi\right|^3}$ and $\partial_{\xi_\gamma}\partial_{\xi_\epsilon}f_i(\bm x_0,\bm \xi)=Gm_i\lb\frac{\delta_{\gamma\epsilon}}{\left|\bm x_0-\bm \xi\right|^3}-\frac{3({x_0}_\gamma-\xi_\gamma)({x_0}_\epsilon- \xi_\epsilon)}{\left|\bm x_0-\bm \xi\right|^5}\rb$.  Therefore we have
\bea
&&\phi_{\rm Newton}(\bm x_0,\bm x^i,\bm y^A)=\phi_{\rm Newton}(\bm x_0,\bm \xi,\bm \zeta)-\sum_i Gm_i\frac{ {x_0}_\gamma-\xi_\gamma}{\left|\bm x_0-\bm \xi\right|^3}{\xi}^i_\gamma-\sum_A Gm_A\frac{ {x_0}_\gamma-\zeta_\gamma}{\left|\bm x_0-\bm \zeta\right|^3}{\zeta}^A_\gamma+\nn
&+&\haf 1\lb\sum_iGm_i\lb \frac{\delta_{\gamma\epsilon}}{\left|\bm x_0-\bm \xi\right|^3}-\frac{3({x_0}_\gamma-\xi_\gamma)({x_0}_\epsilon- \xi_\epsilon)}{\left|\bm x_0-\bm \xi\right|^5}\rb  {\xi}^i_\gamma{\xi}^i_\epsilon  +\right.\nn
&+&\left.\sum_AGm_A\lb\frac{\delta_{\gamma\epsilon}}{\left|\bm x_0-\bm \zeta\right|^3}-\frac{3({x_0}_\gamma-\zeta_\gamma)({x_0}_\epsilon- \zeta_\epsilon)}{\left|\bm x_0-\bm \zeta\right|^5}\rb  {\zeta}^A_\gamma{\zeta}^A_\epsilon        \rb+\cdots 
\eea
Clearly the terms linear in ${\xi}^i_\gamma$ and ${\zeta}^A_\gamma$, the dipole terms vanish simply because
\beq
\sum_im_i{\bm\xi}^i=0\quad \quad\quad \sum_Am_A{\bm\zeta}^A=0.
\eeq
Therefore, we find the quadrupole terms are the first non-trivial terms contributing to the potential. 
\bea
\phi_{\rm Newton}(\bm x_0,\bm x^i,\bm y^A)\ame\phi_{\rm Newton}(\bm x_0,\bm \xi,\bm \zeta)+
\nn
&+&\haf 1\lb\sum_iGm_i\lb \frac{\delta_{\gamma\epsilon}}{\left|\bm x_0-\bm \xi\right|^3}-\frac{3({x_0}_\gamma-\xi_\gamma)({x_0}_\epsilon- \xi_\epsilon)}{\left|\bm x_0-\bm \xi\right|^5}\rb  {\xi}^i_\gamma{\xi}^i_\epsilon  +\right.\nn
&+&\left.\sum_AGm_A\lb\frac{\delta_{\gamma\epsilon}}{\left|\bm x_0-\bm \zeta\right|^3}-\frac{3({x_0}_\gamma-\zeta_\gamma)({x_0}_\epsilon- \zeta_\epsilon)}{\left|\bm x_0-\bm \zeta\right|^5}\rb  {\zeta}^A_\gamma{\zeta}^A_\epsilon        \rb+\cdots \label{dpc}
\eea

However there are other ways of obtaining dipole-like contributions, and we will show that these do not necessarily have to cancel, as exposed in Thorne and Winstein \cite{Thorne:1998hq}.   These contributions are dipole-like, but in fact do not correspond to the total dipole moment of the combined system and the compensatory mass.  The total dipole moment is of course constant or at most linearly time dependent.  
Consider the approximation where all the mass of the system and of the compensatory mass can be taken to be concentrated at their respective centres of mass.  Then, making the assumption that the origin of the coordinate system can be taken such that $|\bm\xi|,|\bm\zeta|\ll |\bm x_0|$ we can expand as
\bea
\phi_{\rm Newton}(\bm x_0,\bm x^i,\bm y^A)&\approx&-\frac{Gm}{\left|\bm x_0 -\bm \xi\right|}-\frac{GM}{\left|\bm x_0-\bm \zeta\right|}
= -G(m+M) \frac{1}{\left|\bm x_0\right|}  -\frac{G\bm x_0}{\left|\bm x_0\right|^3}\cdot\lb{m\bm \xi}+{M \bm \zeta}\rb\nn
\am+ \frac{G}{2\left|\bm x_0\right|^5}\lb m\lb  3(\bm x_0\cdot\bm \zeta)^2  -   \left|\bm \xi\right|^2 \left|\bm x_0\right|^2     \rb +M\lb  3(\bm x_0\cdot\bm \zeta)^2  -   \left|\bm \zeta\right|^2 \left|\bm x_0\right|^2\rb     \rb .\label{mpe}
\eea
However, once again, $m\bm \xi+M\bm \zeta=(m+M)\bm X \ll (m+M)\bm x_0$ and the coordinate system can be chosen so that the centre of mass occurs at $\bm X=0$, giving
\bea
\phi_{\rm Newton}(&\bm x_0&,\bm x^i,\bm y^A)\approx -G(m+M) \frac{1}{\left|\bm x_0\right|}  \nn
\am+ \frac{G}{2\left|\bm x_0\right|^5}\lb m\lb  3(\bm x_0\cdot\bm \zeta)^2  -   \left|\bm \xi\right|^2 \left|\bm x_0\right|^2     \rb +M\lb  3(\bm x_0\cdot\bm \zeta)^2  -   \left|\bm \zeta\right|^2 \left|\bm x_0\right|^2\rb     \rb .\label{mpe2}
\eea
Therefore we see that, in the centre of mass system and if the assumption $|\bm\xi|,|\bm\zeta|\ll |\bm x_0|$ is valid, these dipole terms also must vanish, because of conservation of momentum.  The system may correspond to motions of high accelerations such as jerks, but because of momentum conservation, the counter weight must respond in an identical, compensatory manner, and the combined, total contribution certainly will not exhibit the potentially complicated time dependence of say the system alone. 

\section*{Appendix B :  Detailed Computation for Harmonic Motion}

The detailed computation for the $\hat{x}$ harmonic oscillator is given in this appendix.
We consider the $O_1$ harmonic oscillator (on the left) which motion is described by
\beq
\bm{R}(z) = -\bm{r}(z) = R_0 (1 + \alpha \sin (v z), 0, 0),
\eeq
where $\alpha \equiv \frac{\Delta}{R_0} \ll 1$ is the expansion parameter and $v \equiv \frac{\omega R_0}{c} \ll 1$ is the dimensionless speed such that $v z = \frac{\omega R_0}{c} \cdot \frac{c t_r}{R_0} = \omega t_r$.\\
This motion of a point mass creates a gravitational field, evaluated at the retarded time, is given by Eqn.\eqref{EffectiveGravitoField11} reproduced here
\begin{align}
 \frac{d\bm p}{dt}&= \left. -Gm M \left[ \gamma \frac{\bm{n} - \bm{\beta}}{\kappa^2 R^2} - \frac{\bm{R}}{\kappa^3 R^3} (\dot{R} + \beta^2 - \bm{R}\cdot \dot{\bm{\beta}}) + \frac{4\bm{\beta}}{\kappa^3 R^2} \dot{R} - \frac{4\dot{\bm{\beta}}}{\kappa^2 R} + \frac{\bm{n}}{\kappa^2 R^2} \beta^2 \right] \right|_{ret},
   \label{EffectiveGravitoField1}
\end{align}
where $\kappa = (1 - \bm{\beta} \cdot \bm{n})$ is to be expanded to keep terms of order less than $\beta^3$, $\alpha^3$ and $\bm{n} = \frac{\bm{R}(z)}{R(z)}$. We also have $\bm{n} = \bm{\hat{x}}$, $\bm\beta = -\alpha v \cos(v z) \bm{\hat{x}}$ and $\dot{\bm{\beta}} =\frac{\alpha v^2}{R_0} \sin(v z)  \bm{\hat{x}}$.
It's important to note that, along the computation we are going to neglect the terms proportional to $\beta^3$, $\alpha^4$ and higher.  Then the dynamical equation then is given by
\begin{align}
      \frac{d\bm p}{dt}(z) &= \left. \frac{-Gm M \bm{\hat{x}}}{R_0^2} \left[ \frac{ 1 + \alpha v \cos(vz)  -3 \alpha v^2 \sin(vz) - 3 \alpha^2 v^2 \sin^2(vz) - 3 \alpha^2 v^2 \cos^2(vz)   }{(1+\alpha \sin(vz))^2 (1 + \alpha v \cos(vz))^3 \sqrt{1 - \alpha^2 v^2 \cos^2(vz)}}\right] \right|_{ret}.
\end{align}
To express the field given as a function of $z$, the dimensionless retarded time, as a function $y $, the dimensionless instantaneous time (up to a constant shift) we apply the Lagrange inversion theorem with $g(z)=\frac{dp}{dt}(z)$ where $p$ is the norm of $\bm p$ as all vectors are in one direction :
\begin{align*}
   \frac{dp}{dt}(z) = \frac{dp}{dt} (y) - \al  \sin (vy) \partial_y \frac{dp}{dt} (y) + \frac{\al^2}{2} \partial_y \left(\sin^2 (v y) \partial_y \frac{dp}{dt}(y) \right) + \mathcal{O}(\alpha^3,v^3)
\end{align*}
and with
\begin{align*}
    \partial_y \frac{dp}{dt} (y) &= \frac{-Gm M }{R_0^2} \left(  -2 \frac{\al v \cos (vy)}{(1+\al \sin (vy)  )^3 (1 + \al v \cos (vy) )^2} + 2 \frac{\al v^2 \sin (vy)}{(1+\al \sin (vy))^2}  \right) \\
    \partial_y^2 \frac{dp}{dt}(y) &= \frac{-Gm M }{R_0^2} \left( \frac{6 \al^2 v^2 \cos^2 (vy)}{(1+\al \sin (vy))^4} + 2\frac{\al v^2 \sin (vy)}{(1+\al \sin (vy))^3} \right),
\end{align*}
 where  we have included the second derivative just for completeness, but it does not contribute at the order we are interested in.  Correspondingly all denominators should be expanded to order $v^2$ and $\al^2$. 
After some algebra, we find (suppressing the argument of the trigonometric functions to avoid an unwieldy equation)
\begin{equation*}
     \frac{ -Gm M \bm{\hat{x}}}{R_0^2} \left( \frac{ 1-2\al v \cos -3\al v^2 \sin + 2\al^2 v\cos \sin - 5\al^2v^2 \sin^2 + (1/2)\alpha^2 v^2 \cos^2}{(1+\alpha \sin)^2 } \right) + \mathcal{O}(\al^3, v^3).
\end{equation*}
Every trigonometric function is expressed in terms of $vy = \omega t - v $ however we want our expression to be in terms of $t$ directly. We use a simple trigonometric identities and then Taylor expand up to the second order in $v$,
\begin{align*}
    \sin (v y) &= \sin (\omega t - v) = \sin(\omega t) (1 - v^2/2) - v \cos(\omega t)\\
    \cos (v y) &= \cos (\omega t - v) = \cos (\omega t) + v \sin(\omega t).
\end{align*}
Again we Taylor expand everything and neglect the terms of order $v^3$, $\al^3$ and higher. Then we obtain Eqn.\eqref{egff} in terms  the instantaneous time $t$:
\begin{align}
\frac{d\bm p}{dt} (t)&=  -Gm M \bm{\hat{x}} \left[ \frac{1}{R(t)^2} - \frac{1}{R_0^2} (4\al v^2 \sin(\omega t) + \frac{5}{2}\al^2 v^2 \cos^2(\omega t)  - 4\al^2 v^2 \sin^2(\omega t)  ) \right] \nonumber \\
    &=  -Gm M \bm{\hat{x}}  \left[ \frac{1}{R(t)^2} - \frac{1}{R_0^2} \left(4a(t) + \frac{5}{2}\beta(t)^2 - 4a(t) \Delta(t) \right) \right]
\end{align}

\newpage

\bibliography{DifferenceOfTime.bib}

\begin{thebibliography}{10}

\bibitem{2016LRR....19....1A}
B.~P. {Abbott et. al}.
\newblock {Prospects for Observing and Localizing Gravitational-Wave Transients
  with Advanced LIGO and Advanced Virgo}.
\newblock {\em Living Reviews in Relativity}, 19(1):1, February 2016.

\bibitem{PhysRevD.58.122002}
Scott~A. Hughes and Kip~S. Thorne.
\newblock Seismic gravity-gradient noise in interferometric gravitational-wave
  detectors.
\newblock {\em Phys. Rev. D}, 58:122002, Nov 1998.

\bibitem{Thorne:1998hq}
Kip~S. Thorne and Carolee~J. Winstein.
\newblock {Human gravity gradient noise in interferometric gravitational wave
  detectors}.
\newblock {\em Phys. Rev. D}, 60:082001, 1999.

\bibitem{ParanjapeSpeedGravity}
MB~Paranjape.
\newblock How to measure the speed of gravity.
\newblock {\em arXiv preprint arXiv:1208.2293}, 2012.

\bibitem{LIGOSpeedGravity}
BP~Abbott, LIGO~Scientific Collaboration, Virgo Collaboration, M~Sakellariadou,
  and Fermi Gamma-ray Burst.
\newblock Gravitational waves and gamma-rays from a binary neutron star merger:
  Gw170817 and grb 170817a.
\newblock {\em The Astrophysical Journal Letters}, 848(L13):27pp, 2017.

\bibitem{PhysRevLett.119.161102}
Neil Cornish, Diego Blas, and Germano Nardini.
\newblock Bounding the speed of gravity with gravitational wave observations.
\newblock {\em Phys. Rev. Lett.}, 119:161102, Oct 2017.

\bibitem{Saulson:2017jlf}
Peter~R. Saulson.
\newblock {\em {Fundamentals of Interferometric Gravitational Wave Detectors}}.
\newblock World Scientific, 2nd. ed. edition, 2017.

\bibitem{Carroll}
Sean~M Carroll.
\newblock {\em Spacetime and geometry}.
\newblock Cambridge University Press, 2019.

\bibitem{Misner}
Charles~W Misner, Kip~S Thorne, and John~Archibald Wheeler.
\newblock {\em Gravitation}.
\newblock Princeton University Press, 2017.

\bibitem{bc}
Ben Craven.
\newblock Kinetic energy of a drifting plate:
  http://bencraven.org.uk/2017/05/22/the-kinetic-energy-of-a-drifting-tectonic-plate/.

\bibitem{Bletery1027}
Quentin Bletery, Amanda~M. Thomas, Alan~W. Rempel, Leif Karlstrom, Anthony
  Sladen, and Louis De~Barros.
\newblock Mega-earthquakes rupture flat megathrusts.
\newblock {\em Science}, 354(6315):1027--1031, 2016.

\bibitem{Weinberg}
Steven Weinberg.
\newblock Gravitation and cosmology: principles and applications of the general
  theory of relativity.
\newblock 1972.

\bibitem{Ryder}
Lewis Ryder.
\newblock {\em Introduction to general relativity}.
\newblock Cambridge University Press, 2009.

\bibitem{Jackson:1998nia}
John~David Jackson.
\newblock {\em {Classical Electrodynamics}}.
\newblock Wiley, 1998.

\bibitem{Carlip}
Steve Carlip.
\newblock Aberration and the speed of gravity.
\newblock {\em Physics Letters A}, 267(2-3):81--87, 2000.

\bibitem{Kinnersley:1969zz}
William Kinnersley.
\newblock {Field of an Arbitrarily Accelerating Point Mass}.
\newblock {\em Phys. Rev.}, 186:1335--1336, 1969.

\bibitem{Aichelburg:1971xy}
P.~C. Aichelburg and R.~U. Sexl.
\newblock On the gravitational field of a massless particle.
\newblock {\em General Relativity and Gravitation}, 2(4):303--312, 1971.

\bibitem{Martynov2016}
Denis~V Martynov, ED~Hall, BP~Abbott, R~Abbott, TD~Abbott, C~Adams,
  RX~Adhikari, RA~Anderson, SB~Anderson, K~Arai, et~al.
\newblock Sensitivity of the advanced ligo detectors at the beginning of
  gravitational wave astronomy.
\newblock {\em Physical Review D}, 93(11):112004, 2016.

\bibitem{laplace1799traite}
P.S. Laplace and J.B.M. Duprat.
\newblock {\em Trait{\'e} de m{\'e}canique c{\'e}leste}.
\newblock Trait{\'e} de m{\'e}canique c{\'e}leste /par P.S. Laplace ... ; tome
  premier [-quatrieme]. de l'Imprimerie de Crapelet, 1799.

\bibitem{VanFlandern}
Tom Van~Flandern.
\newblock The speed of gravity---what the experiments say.
\newblock {\em Physics Letters A}, 250(1-3):1--11, 1998.

\bibitem{Will:2003yj}
Clifford~M. Will.
\newblock {Propagation speed of gravity and the relativistic time delay}.
\newblock {\em Astrophys. J.}, 590:683--690, 2003.

\bibitem{Lagrange1770}
Joseph~Louis Lagrange.
\newblock {\em Nouvelle m{\'e}thode pour r{\'e}soudre les probl{\`e}mes
  ind{\'e}termin{\'e}s en nombres entiers}.
\newblock Chez Haude et Spener, Libraires de la Cour \& de l'Acad{\'e}mie
  royale, 1770.

\bibitem{Whittaker}
Edmund~Taylor Whittaker and George~Neville Watson.
\newblock {\em A course of modern analysis}.
\newblock Cambridge university press, 1996.

\bibitem{Parikh:2020nrd}
Maulik Parikh, Frank Wilczek, and George Zahariade.
\newblock {The Noise of Gravitons}.
\newblock {\em Int. J. Mod. Phys. D}, 29(14):2042001, 2020.

\bibitem{parikh2020signatures}
Maulik Parikh, Frank Wilczek, and George Zahariade.
\newblock Signatures of the quantization of gravity at gravitational wave
  detectors.
\newblock arXiv: hep-th 2010.08208, 2020.

\bibitem{HarmsJ2015Tgpi}
J~Harms, J.-P Ampuero, M~Barsuglia, E~Chassande-Mottin, J.-P Montagner, S.~N
  Somala, and B.~F Whiting.
\newblock Transient gravity perturbations induced by earthquake rupture.
\newblock {\em Geophysical journal international}, 201(3):1416--1425, 2015.

\bibitem{cdi_oup_primary_10_1093_gji_ggaa486}
Tomofumi Shimoda, K{\'e}vin Juhel, Jean-Paul Ampuero, Jean-Paul Montagner, and
  Matteo Barsuglia.
\newblock Early earthquake detection capabilities of different types of
  future-generation gravity gradiometers.
\newblock {\em Geophysical journal international}, 224(1):533--542.

\bibitem{JuhelK2018EEWU}
K~Juhel, J.~P Ampuero, M~Barsuglia, P~Bernard, E~Chassande-Mottin, D~Fiorucci,
  J~Harms, J.P. Montagner, M~Vall{\'e}e, and B.~F Whiting.
\newblock Earthquake early warning using future generation gravity
  strainmeters.
\newblock {\em Journal of geophysical research. Solid earth},
  123(12):10,889--10,902, 2018.

\bibitem{ww}
Martin Vall{\'e}e, Jean~Paul Ampuero, K{\'e}vin Juhel, Pascal Bernard,
  Jean-Paul Montagner, and Matteo Barsuglia.
\newblock Observations and modeling of the elastogravity signals preceding
  direct seismic waves.
\newblock {\em Science (American Association for the Advancement of Science)},
  358(6367):1164--1168, 2017.

\bibitem{SCHELLART201341}
W.P. Schellart and N.~Rawlinson.
\newblock Global correlations between maximum magnitudes of subduction zone
  interface thrust earthquakes and physical parameters of subduction zones.
\newblock {\em Physics of the Earth and Planetary Interiors}, 225:41--67, 2013.

\bibitem{nz}
GNS Science.
\newblock How long do earthquakes last:
  https://www.gns.cri.nz/home/learning/science-topics/earthquakes/monitoring-earthquakes/other-earthquake-questions/how-long-does-an-earthquake-last.

\end{thebibliography}
\bibliographystyle{unsrt}

\end{document}